\documentclass[a4paper,11pt,dvips]{article}
\usepackage{graphicx}
\usepackage{amsmath}
\usepackage{amssymb}
\usepackage{latexsym}
\usepackage{color}
\usepackage{cite}
\usepackage{makeidx}

\newcommand{\bra}{\begin{array}}
\newcommand{\era}{\end{array}}
\newcommand{\beq}{\begin{equation}}
\newcommand{\eeq}{\end{equation}}
\newcommand{\bqr}{\begin{eqnarray}}
\newcommand{\eqr}{\end{eqnarray}}

\def\BC{\bb C}
\def\_\BC{\bbi C}

\def\( {\left(}
   \def\) {\right)}
\def\[ {\left[}
\def\] {\right]}
\def\no2 {{\textstyle{n\over 2}}}

\newcommand{\om}{\omega}

\newcommand{\si}{\sigma}

\newcommand{\te}{\theta}

\newcommand{\pa}{\partial}

\newcommand{\lga}{\longrightarrow}

\newcommand{\da}{\dagger}

\newcommand{\ov}{\over}

\newcommand{\sq}{\sqrt}

\newcommand{\ev}{\equiv}
\newcommand{\lb}{\label}

\begin{document}
\begin{titlepage}
\setcounter{page}{1}
\renewcommand{\thefootnote}{\fnsymbol{footnote}}

\begin{flushright}
\end{flushright}

\vspace{5mm}

\begin{center}
 {\Large \bf 
 Hall Conductivities for Confined System in Noncommutative Plane }\\

\vspace{5mm}

{\bf Kamal El Asli}$^{a}$, {\bf Rachid Hou\c{c}a}$^{a}$ and {\bf Ahmed Jellal\footnote{\sf
ajellal@ictp.it -- a.jellal@ucd.ac.ma}}$^{a,c,d}$

\vspace{5mm}

{$^{a}$\em Theoretical Physics Group,  
Faculty of Sciences, Choua\"ib Doukkali University,\\
24000 El Jadida, Morocco}

{$^b$\em Saudi Center for Theoretical Physics, Dhahran, Saudi
Arabia}

{$^c$\em Physics Department, College of Sciences, King Faisal University,\\
Alahsa 31982, Saudi Arabia}

\vspace{3cm}

\begin{abstract}

We propose an approach based on the generalized quantum mechanics to deal with the basic features of the spin Hall effect.
We 
begin by
considering two decoupled harmonic oscillators on the noncommutative plane and
determine the solutions of the energy spectrum. We realize two algebras in terms of
the quadratic observables and show their importance in filling the shells with
fermions.
Under some transformation we show that our system is submitted to an effective  Lorentz force similar to
that acting on one particle in an external magnetic field. From equation of motions, we end up with
the charge and spin Hall conductivities as function of the noncommutative parameter $\te$. By switching off $\te$
we recover standard results developed on the subject and in the limit $\te\lga 0$ we show that
our approach can reproduce the Laughin wavefunctions.

\end{abstract}
\end{center}
\vspace{3cm}



\end{titlepage}

\section{Introduction}

In 1971 D'yakonov and Perel predicted from a phenomenological model that spin-orbit
coupling should lead to a new family of Hall effects~\cite{yakonov,yakonov1}, known
actually
as the spin Hall effect (SHE).
It is a consequence of an applied electric field to a sample
that leads to a spin transport in perpendicular direction and spin accumulation at the lateral edges~\cite{kato,wunderlich}.
Its most remarkable feature  is a 2D semiconductor subject to a difference of the potential, which can split the spin to up and down
components. The spin current  produces
an accumulation of spin-up electrons on one side of the excited region and spin-down electrons on the other side, with no total carrier accumulation on either side.
SHE is characterized by a spin Hall conductivity resulting from the spin polarization on the boundaries
of the sample.
Spin-orbit effects
are generally classified as extrinsic effects because
they arise from scattering of impurity potentials. Another
class of spin-orbit effects, known as intrinsic effects, is a consequence of the inherent band structure of a crystalline material.
These effects can lead to a new state of matter called the quantum SHE, which has been predicted
theoretically in 2000 by the Kane-Mele model \cite{Kane} and the experimental realization came later on~\cite{markus}.
 The innovative aspect of the  quantum SHE is that it appears in the absence of magnetic field
 and then there is no symmetry breaking of time reversal.

On the other hand, the noncommutative geometry~\cite{connes} plays  an important role in physics
and was used to solve many issues in different areas. 
For instance, 
interesting results were reported for the quantum Hall effect~\cite{Prange} due either to the charge current~\cite{jellal}
or spin current~\cite{houca,dayi,ma,basu}.
To remember, the noncommutative geometry 
is already exits and found its application in the fractional quantum Hall effect when the lowest Landau Level (LLL) is partially filled.
It happened that in
 LLL, the potential energy is strong enough
than kinetic energy and therefore the particles are glue in the fundamental level.
%
As a consequence of
this drastic reduction of the degrees of freedom, the two space coordinates  become noncommuting~\cite{jellal2}
and
satisfy the commutation relations analogue to those verifying by the position and the momentum in quantum
mechanics. In this case, the electron is not a point like particle anymore and can at best
be localized at the scale of the magnetic length.
In the present work, we serve form the mentioned mathematical tool to expose our idea and thus offer
another way to study SHE for a confined system in two dimensions.

 Based on the 
 results presented in~\cite{houca,Chudnovsky}, 
we quantum mechanically develop an approach to study SHE for
particles living on the noncommutative plane $\mathbb{R}_{\theta}^2$.
For this, we use the star product to define the Hamiltonian system that captures the basic features
of the effect and allows for building an effective theory.
This Hamiltonian is nothing but that describing two 
coupled harmonic oscillators where the solutions of the energy spectrum are algebraically derived
as functions of the noncommutative parameter $\theta$.
Different interpretations are presented  in order to  give
some experimental evidence of $\te$ and show its relevance in modern physics. 
After realizing the algebras $su(2)$ and $su(1,1)$, we discuss the possibilities
how to fill
the shells with fermions.
Using the Hamilton canonically equations to get the average velocity of particles, we explicitly determine
the 
charge  and  spin Hall conductivities
in terms  of $\theta$. 
 The ratio between these two conductivities shows an independency of $\te$ and the concentration of charge carriers, which is in agreement with the results
 obtained in
 \cite{Chudnovsky}.
 Finally interesting cases are examined and in particular the limiting case $\te\lga 0$ is analyzed to show that $\te$ can be
 identified to an external magnetic field and therefore make contact with Laughin states~\cite{laughlin}.

The present paper is organized as follows. In section 2, we consider two decoupled harmonic oscillators on the noncommutative plane $\mathbb{R}_{\theta}^2$.
This process allows us to end up with a Hamiltonian system similar to that of one particle living on the plane
in the presence of an external magnetic field, known as the Landau problem. In section 3, we introduce the annihilation
and creation operators to easily
determine the eigenvalues and eigenstates. In section 4, we show that there are two
algebras those can be used in filling the shells with fermions.
We define an effective force as function of $\te$ similar to the Lorentz force
acting on one particle in section 5. By establishing the equations of motions describing
our system we use the current definition to obtain
the Hall conductivities
in terms of $\theta$ in section 6. By fixing $\theta$, we show that it is possible
to recover interesting results dealing with  SHE as well as quantum Hall effect in section 7. We conclude our work in the final section.

\section{Two harmonic oscillators in noncommutative space} 

We consider two decoupled harmonic oscillators on the noncommutative
plane $\mathbb{R}^2_{\te}$
and determine the corresponding eigenvalues as well as
eigenstates.  In doing so,
we introduce the star product and the ordinary commutation relations in
quantum mechanics.
Before doing this, we recall the algebraic structures of the ordinary
system on the plane $\mathbb{R}^2$ and the  solution of the energy spectrum.


As we claimed before our proposal can be elaborated by considering two decoupled harmonic oscillators
having the same masses $m$ and frequencies $\om$. These are described by the Hamiltonian
\beq\lb{ccc}
H={1\over 2m}
\left(p_{x}^{2 }+p_{y}^{2}\right) +{m\om^2\over 2}\left(x^2+y^2\right)
\eeq
which is also the Hamiltonian describing single particle living on $\mathbb{R}^2$  and subjected to
a confining potential. 
It can be diagonalized by introducing the creation and annihilation operators
\beq
b_i= {1\over\sqrt{2\hbar m\om}}p_i -i \sqrt{m\om\over 2\hbar},
\qquad b_i^{\da}= {1\over\sqrt{2\hbar m\om}}p_i +i \sqrt{m\om\over 2\hbar}
\eeq
where  $i=x,y$ and the only non-vanishing commutator are
\beq
\left[b_i, b_i^{\da} \right]= \mathbb{I}.
\eeq
Now we can use these operators to map the Hamiltonian \eqref{ccc} as 
\beq
H={\hbar \om \over 2}\left(b^{\da}_x b_x + b^{\da}_y b_y+1\right).
\eeq
Clearly from the eigenvalue equation $H |n_x,n_y\rangle =  E_{n_x,n_y} |n_x,n_y\rangle$, one can easily get
the eigenstates
\beq
|n_x,n_y\rangle= {(b^{\dagger}_x)^{n_x}\over
  \sq{n_x!}}{(b^{\dagger}_y)^{n_y}\over \sq{n_y!}}\ |0,0\rangle
\eeq
as well as the eigenvalues
\beq\lb{sssp}
E_{n_x,n_y}={\hbar \om \over 2}\left(n_x+n_y+1\right), \qquad n_i = 0,1,2,\cdots
\eeq
where
$|0,0\rangle$ is the fundamental state.
Next, we will see how these results will be generalized for
one particle living on
$\mathbb{R}^2_{\te}$. 

Our main goal is to investigate  the basic features of particles
living on the noncommutative plane and in particular study SHE. To do this task, we need
to settle a theoretical model that allows us to shed light
on different issues. 
We adopt a method similar to
that used in~\cite{jellal} where
the canonical quantization of the system described by the Hamiltonian (\ref{ccc})
is achieved by
introducing the coordinate $r_j$ and momentum $p_k$ operators
satisfying the commutation relation
\beq \lb{ddd}
\left[r_j,p_k\right]=i\hbar \delta_{jk}.
\eeq
But to deal with our proposal,  we consider a generalized
quantum mechanics governed by
(\ref{ddd}) and the noncommutative coordinates
\beq \lb{aaa}
\left[x,y\right]=i\te
\eeq
where $\te$ is a real free parameter and has length square
of dimension.
Without loss of generality, hereafter we assume that $\te>0$ is fulfilled.
Noncommutativity can be imposed
by treating the coordinates as commuting but requiring that
composition of their functions is given in terms of the star
product
\beq \lb{abc}
\star \equiv \exp{i\te\over 2}\left(\overleftarrow {\pa_x}\overrightarrow{\pa_y}-
\overleftarrow{\pa_y}\overrightarrow{\pa_x}\right).
\eeq
Now, we deal with the commutative coordinates $x$ and $y$ but replace the
ordinary products with the star product~(\ref{abc}). For example, instead of the
commutator~(\ref{aaa}) one defines
\beq
x\star y-y\star x=i\te.
\eeq
At this level,
let us  derive the corresponding form of the Hamiltonian
(\ref{ccc}) in terms of the noncommutative
coordinates~(\ref{aaa}).
First, we quantize the present system by establishing the
commutation relation~(\ref{ddd}). Second, we take into account
the noncommutativity of the coordinates
by defining a new Hamiltonian operator as
\beq
H\star \psi(\vec {r})\equiv H^{\sf nc}\psi(\vec {r})
\eeq
where $\psi(\vec {r})$ is an arbitrary eigenfunction of $H$.

It is clear now that from the above materials we can easily
write the noncommutative version of
the Hamiltonian (\ref{ccc}). This is
%
\beq\lb{rrr}
H^{\sf nc}={1\ov2m_{\theta}} \left(p_x^{2}+p_y^{2}\right)
+{m\omega^{2}\over{2}}{\left(x^{2}+y^{2}\right)+{m\omega^{2}\over2\hbar}\te\left(yp_x-xp_y\right)}
\eeq
where the effective mass is given by
\beq\lb{m_te}
m_\te= \frac{m}{1+\left(\frac{m\om\te}{2\hbar}\right)^2}
\eeq
This form of mass suggest an interesting way to measure the noncommutative parameter and give
hint ont its values. In doing so, one can make comparison with a relativistic particle of mass
$m_0$ and moving with a velocity $v$, which acquires an effective mass
\beq
m_{\sf eff}=\frac{m_0}{\sqrt{1-\frac{v^2}{c^2}}}.
\eeq
Now by identifying both masses and requiring $m=m_0$, one can immediately fix $\te$ in terms of speed of light and $v$ as
\beq
\te_\pm = \pm\frac{2\hbar}{m\om} \left(-1+\sqrt{1-\frac{v^2}{c^2}}\right)^{1/2}
\eeq
We have some remarks in order. Indeed, we emphasis that
due to the noncommutativity between spacial coordinates we ended up with
two coupled harmonic oscillators \eqref{rrr} where the coupling is described by
 the angular momenta $L_z(\te)$ term
\beq\lb{amo}
L_z(\te) ={m\om^2\te\over 2\hbar} \left(yp_x-xp_y\right)
\eeq
and then we can split $H^{\sf nc}$ into free part and  $L_z(\te)$
\beq\lb{sl2p}
H^{\sf nc}= H^{\sf nc}_0 + L_z(\te)
\eeq
It is obvious that (\ref{amo}) disappears once we set $\te=0$ and then we
recover  (\ref{ccc}). We notice that $H^{\sf nc}$ is sharing some common features
with the Hamiltonian describing one particle moving in plane and subjected to an external magnetic field.
This statement will be clarified later on and will give another interesting feature of
$\te$.

\section{Solution of the energy spectrum}

In order to get
the eigenvalues and eigenstates for our
system, we proceed by making use of the algebraic method to diagonalize the Hamiltonian \eqref{rrr}.
For this, we introduce a pair of the annihilation and creation operators
\bqr
&& a_x= {x\over\l_\te}+i{l_\te\over 2\hbar}p_x, \qquad a_x^{\da}= {x\over\l_\te}-i{l_0\over 2\hbar}p_x\\
&& a_y={y\over\l_\te}+i{l_\te\over 2\hbar}p_y, \qquad a_y^{\da}={y\over\l_\te}-i{l_0\over 2\hbar}p_y
\eqr
where we have defined the noncommutative length as
\beq
l_\te={\sqrt[4]{ \left(\frac{2{\hbar}}{m\om}\right)^2+{\te}^{2}}}. 
\eeq
Actually, this can be compared to the magnetic length  $l_B=\sqrt{\frac{e\hbar}{cB}}$ in order
to extract more information about our system, we will return
to discuss this point later.
One can easily show the commutation relations
\beq
\left[a_x, a_x^{\da} \right]=\left[a_y, a_y^{\da} \right]= \mathbb{I}.
\eeq
Now let us set a new pair of the shell  operators
\bqr
&& a_d={ 1\over\sqrt{2}}({a_x-ia_y}), \qquad a_d^\da={ 1\over\sqrt{2}}({a_x^\da+ia_y^\da})\\
&& a_g={ 1\over\sqrt{2}}(a_x+ia_y), \qquad a_g^\da={ 1\over\sqrt{2}}(a_x^\da-ia_y^\da)
\eqr
verifying
\beq
\left[a_d, a_d^{\da} \right]=\left[a_g, a_g^{\da} \right]= \mathbb{I}
\eeq
and all other commutators are vanishing. After some algebra, we can
map the phase space coordinates in terms of the above operators as
\bqr
x= {l_0\over{2\sqrt{2}}} ({a_d+a_d^{\da}+a_g+a_g^{\da}}),
\qquad p_x={\hbar\over{\sqrt{2}il_0}}(a_d-a_d^{\da}+a_g-a_g^{\da})\\
y={l_0\over{2\sqrt{2}i}}(a_d^{\da}-a_d+a_g-a_g^{\da}),
\qquad p_y={\hbar\over{\sqrt{2}l_0}}(a_d+a_d^{\da}-a_g-a_g^{\da}).
\eqr
Combining all as well as introducing
 the operator numbers $N_d=a_d^{\da}a_d$ and
$N_g=a_g^{\da}a_g$, we show that the Hamiltonian \eqref{rrr} takes the form
\beq\lb{ncham}
H^{\sf nc}={m\omega^{2}\over{2}}(l_\te^{2}+\te)N_d+{m\omega^{2}\over{2}}(l_\te^{2}-\te)N_g+{m\omega^{2}\over{2}}{l_\te^{2}}.
\eeq

In order to determine the solution of the energy spectrum corresponding to \eqref{ncham}, we
solve the eigenvalue equation
\beq
H^{\sf nc}|n_d,n_g\rangle=E_{n_d,n_g}|n_d,n_g\rangle
\eeq
to end up with the eigenvalues
\beq\lb{rerr}
E_{n_d,n_g}= {m\omega^{2}l_\te^2\over{2}}(n_d+n_g +1) + {m\omega^{2}\te\over{2}}(n_d - n_g)
\eeq
and  eigenstates
\beq
|n_d,n_g\rangle= {(a^{\dagger}_d)^{n_d}(a^{\dagger}_g)^{n_g}\over
  \sq{(n_d!)(n_g!)}} |0,0\rangle
\eeq
where  the quantum numbers $n_d$ and $n_g$ are non-negative integers.
At this stage, we have two remarks in order. Firstly
after making comparison, one can interpret the second part in  \eqref{rerr}
as a quantum correction to decoupled spectrum $E_{n_x,n_y}$ \eqref{sssp}.
Secondly in the limit $\te\lga 0$, \eqref{rerr} reduces to that of one particle
in an external magnetic field $B$ corresponding to the symmetric gauge where $\te$ will play the role
of  $B$.

\section{Algebras and filling shells}

Based on the work \cite{gazeau}, we show that the noncommutativity of our system 
allows us to realize 
two distinct dynamical symmetries 
in terms of the  quadratic observables. Indeed, the generators of the algebra
${su}(2)$ can be constructed as
\beq
S_+ = a_d^{\dagger} a_g, \qquad
S_- = a_g^{\dagger} a_d, \qquad
S_z = \frac{N_d - N_g}{2} = \frac{L_z}{2\hbar}
\eeq
which satisfy the $su(2)$ commutation relations
\beq
\lbrack S_+,S_- \rbrack = 2S_z,  \qquad
\lbrack S_z,S_{\pm} \rbrack = \pm S_{\pm},
\eeq
and the invariant Casimir operator is given by
\beq
{\cal C} 
         = \left(\frac{N_d + N_g}{2}\right)\left(\frac{N_d + N_g}{2} + 1\right).
\eeq
Therefore, for a fixed value $j = (n_d+n_g)/2$ of the operator
$(N_d + N_g)/2 = {\cal H}_0^{\sf nc}/(m\om l_\te^2) - 1/2$,
there exists a $(2j + 1)$-dimensional unitary irreducible representation (UIR) of ${su}(2)$ in which the
operator $ S_z$ has its spectral values in the range
$ -j \leq k \leq j$, with $k = (n_d - n_g)/2$.

Also we have another free room for a second algebra that is ${su}(1,1)$. The associated
generators can be defined as
\beq
T_+ = a_d^{\dagger} a_g^{\dagger}, \qquad
T_- = a_g a_d, \qquad
T_0 = 
\frac{{\cal H}_0^{\sf nc}} {m\om l_\te^2}
\eeq
showing the algebra
\beq
\lbrack T_+,T_- \rbrack = -2T_0, \qquad
\lbrack T_0,T_{\pm} \rbrack = \pm T_{\pm}.
\eeq
and the Casimir operator is given by
\beq
{\cal D} 
        =- \left(\frac{L_z^2(\te)}{(m\om^2 \te)^2} -\frac{1}{4}\right).
\eeq
There are two  interesting situation one has to consider. First when $n_d \geq n_g$,
for a fixed value $k +1/2 \geq 1/2 $ of the operator
${L_z(\te)}/({m\om^2 \te}) +1/2$, there exists a UIR of ${su}(1,1)$ in the discrete series, in
which the operator $ T_0 $ has its spectral values in the infinite range
$ k +1/2, k +3/2, k + 5/2, \cdots$.
Second when  $n_d \leq n_g$, for a fixed value
$-k+1/2 \geq 1/2$ of the operator ${-L_z(\te)}/({m\om^2 \te}) +1/2$,
there also exists a UIR of ${su}(1,1)$ in which the spectral value of the
operator $ T_0 $ runs in the infinite
range $ -k +1/2, -k +3/2, -k + 3/2, \cdots$.


According to the above two algebras, we can offer
a way how to order the quantum numbers $n_d$ and $n_g$.
In what follows, we consider three interesting cases where
in the week and strong limits we will see the manifestation
of $su(2)$ and $su(1,1)$, respectively. In the generic case,
we show under circumstance it is possible to have degeneracy
of filling the levels.\\
$\blacktriangleright$ {For the week case} we assume that
the product fulfilled $\om\te \ll 1$.
Let us write the eigenvalues in terms of the algebra ${su}(2)$, such as
\beq
E_{n_d, n_g}\ev E_{j,k}= m\om^2 l^2_\te j+ m\om^2 \te k +\frac{1}{2} m\om^2l^2_\te
\eeq
Then in the present case it reduces as
\beq
E_{j} \approx \hbar \om (2j + 1) 
\eeq
which is actually $\te$-independent. In the present case,
${su}(2)$  becomes a true
symmetry of the Hamiltonian, which explains the degeneracy of order
$2j + 1$ for the level $E_{j}$.
Note that there are $(j_0 + 1) (2j_0 + 1)$ (spinless) electrons
which fill the shells up to $j_0$.\\
$\blacktriangleright$ {For strong case, we consider the opposite limit, i.e. $\om\te \gg 1$. Therefore the energy
can be approximated as
\beq
E_{n_d } \approx {m\om^2 \te} \left(n_d + \frac{1}{2}\right).
\eeq
which tells us, for a given value of $n_d$, that there is an infinite degeneracy labeled
by $n_g$ or by $2k \leq n_d$ 
where the quantum number $n_d$ corresponds to the Landau level index.
The present situation can be linked with the algebra ${su}(1,1)$ by claiming that 
for a given value of $k \leq 0$,
the energy eigenstates are ladder states for the discrete series
representation labeled by $ -k + 1/2$.\\
$\blacktriangleright$ In the uncommensurate generic intermediate case, which means $(l_\te^2 +\te)/(l_\te^2 -\te) \notin \mathbb{Q}$
and no approximation is relevant,
we are faced to the problem of ordering the relatively dense (but not uniformly
discrete) set of eigenvalues
\beq
{\cal E}_{n_d, n_g} \equiv \frac{E_{n_d, n_g}}{m\om^2(l_\te^2 -\te)/2}-\frac{l_\te^2}{l_\te^2 -\te}
                     = \frac{l_\te^2 +\te}{l_\te^2 -\te}n_d + n_g.
\eeq
However, in the commensurate case, $(l_\te^2 +\te)/(l_\te^2 -\te)  = p/q \in \mathbb{Q}$, degeneracy can
be found if one requires the condition
\beq
\frac{p}{q} = - \frac{n_g - n'_g}{n_d - n'_d}.
\eeq
and therefore in this situation we have $E_{n_d, n_g} = E_{n'_d, n'_g}.$

\section{Effective Lorentz force}

Based on the approach developed in\cite{Chudnovsky},
we begin by deriving an affective force
analogue to that of Lorentz
acting on one particle of charge $e$ in
an external magnetic field.
This will be helpful in sense that one can solve different issues
and in particular those related to the Hall effect. As it will be clear later on,
one can turn out $te$ to recover interesting results developed on
the subject.

To fix our ideas, let us
associate our confining potential $V(\vec{r})=\frac{m\om^2}{2} r^2$ to
an electric field $E$  through the usual relation
$\vec{E}=-{1\ov e}\vec\nabla V(\vec r)$, with the vector position $\vec r=(x,y)$.
This mapping can be used to write the Hamiltonian (\ref{rrr}) as
\beq\lb{H1}
H^{\sf nc}={\vec{p}^{2}\ov2m_{\theta}} +{e\over{2\hbar}}\vec{\theta} \cdot {\left(\vec{E}\times\vec{p}\right)}+V\left(\vec{r}\right)
\eeq
where we have defined a noncommutative vector $\vec{\te}$ as follows
\beq
\vec{\te}=
\left(
  \begin{array}{c}
    \te_x \\
    \te_y \\
    \te_z\\
  \end{array}
\right).
\eeq
We notice that the second term in \eqref{H1} is nothing but the scalar product between $\vec \te$ and the angular momenta
$\vec L= \vec r \times \vec p$. Then by recalling the spin-orbit coupling $\vec S \cdot \vec L$, one can
give another way to measure the noncommutative parameter in terms of the spin under the request
$\te_x\equiv S_x$, $\te_y\equiv S_y$ and $\te_z\equiv S_z$. This can be generalized to describe
interesting features of the spin-orbit coupling and therefore
make contact with the Rashba and Dresselhaus couplings~\cite{rashba}, which are the cornerstones of
different spin Hall effects.

Now let us examine the dynamical behavior of our system and write the corresponding equations of motions. Indeed,
from \eqref{H1}
we show that the
Hamiltonian mechanics for canonically conjugated variables $\vec{p}$ and $\vec{r}$ is governed by the equations
\begin{eqnarray}\lb{cnju}
\dot{\vec{r}}&=& 
{\vec{p}\ov m_\theta}+{e\ov2\hbar}\left[\vec{\theta}\times \vec{E}\right] \lb{cnju}\\ 
\dot{\vec{p}}&=& 
-{\partial V(\vec{r})\ov \partial \vec{r}}-{e\ov2\hbar}{\partial\ov \partial \vec{r}}\left(\left[\vec{\theta}\times \vec{E}\right]\cdot \vec{p}\right)\lb{cnju2}
\end{eqnarray}
leading to the relations
\begin{eqnarray}\lb{frdr}
\vec{p}&=&m_\theta\dot{\vec{r}}-{em_\theta \ov2\hbar}\left[\vec{\theta}\times \vec{E}\right]\\
\dot{\vec{p}}&=&m_\theta\ddot{\vec{r}}-{em_\theta \ov2\hbar}\left(\dot{\vec{r}}{\partial\ov \partial \vec{r}}\right)\left[\vec{\theta}\times \vec{E}\right].
\end{eqnarray}
After substituting these   into (\ref{cnju}) and \eqref{cnju2}, we end up with the following form of the second Newton's law for charge carriers
\begin{eqnarray}\lb{gtgt}
m_\theta\ddot{\vec{r}}=-{\partial V(\vec{r})\ov \partial \vec{r}}+{em_\theta\ov2\hbar}\left\{\left(\dot{\vec{r}}{\partial\ov \partial \vec{r}}\right)\left[\vec{\theta}\times \vec{E}\right]-{\partial\ov \partial \vec{r}}\left(\left[\vec{\theta}\times \vec{E}\right]\cdot \dot{\vec{r}}\right)\right\} + \frac{e^2 m_\te}{4\hbar} {\partial\ov \partial \vec{r}} \left[\vec\te \times \vec E\right]^2.
\end{eqnarray}
To simplify our task and proceed further, let us introduce a convenient
approximation. Indeed, we neglect
the term proportional to $\left[\vec{\theta}\times \vec{E}\right]^2$ and then write the last equation as
\beq\lb{drag}
m_\theta\ddot{\vec{r}} \simeq -{\partial V(\vec{r})\ov \partial \vec{r}} +
\vec{F}(\vec{\theta},\vec{r},\dot{\vec{r}})
\eeq
where we have fixed $\vec F$ as
\begin{eqnarray}
\vec{F}(\vec{\theta},\vec{r},\dot{\vec{r}})
= -{em_\theta\ov2\hbar}\dot{\vec{r}}\times\left({\partial\ov \partial \vec{r}}\times\left[\vec{\theta}\times \vec{E}\right]\right)
\end{eqnarray}
Now if we look at the shape of the Lorentz force, one can immediately conclude
that $\vec F$ is its analogue and can be mapped into
\beq\lb{force}
\vec{F}(\vec{\theta},\vec{r},\dot{\vec{r}})={e\ov c}\left(\dot{\vec{r}}\times \vec{B}(\vec{\theta})\right)
\eeq
where the effective magnetic field can be defined in terms of the vector potential $\vec{A}(\vec{\theta})$
via standard relation $\vec{B}(\vec{\theta})=\vec{\nabla}\times \vec{A}(\vec{\theta})$,
such as
\beq\lb{A}
\vec{A}(\vec{\theta})={cm_\theta \ov2\hbar}\left[\vec{\theta}\times \vec{E}\right].
\eeq

Keeping in mind the Hamiltonian structures for one particle
in magnetic field, known as Landau problem, we can then use the above tool to write
(\ref{H1}) in a compact form. Indeed,
after neglecting the square term in $\te$
we find
\beq\lb{HHH}
H^{\sf nc} \simeq {1\ov2m_{\theta_z}}\left(\vec{p}+{e\ov c}\vec{A}(\vec{\theta})\right)^2+V(\vec{r}).
\eeq
It is clearly seen that this Hamiltonian can be interpreted as 
one describing
a particle living on the plane in the presence of the effective magnetic field $\vec{B}(\vec{\theta})$
and scalar potential $V(\vec{r})$.
Based on our knowledge, it turns out that \eqref{HHH} can be used as model
to build an effective theory for
the  fractional quantum Hall effect~\cite{Prange} and related
matters.

Now  let us return to discus the product $\vec \te \times \vec E$ and see what we can extract as
information regarding our system. To fix our ideas,
we choose hereafter the noncommutative components as $\te_x=\te_y=0$ and $\te_z=\te$. From the confining
potential we can easily obtain the electric field
\beq
 \vec E= -\frac{m\om^2}{e}(x,y)
\eeq
  and then \eqref{A}
gives
\beq\lb{AA}
\vec A(\vec\te) = 
\frac{2\hbar c\te/e}{\left(\frac{2\hbar}{m\om}\right)^2 +\te^2} (y,-x).
\eeq
It can be compared to the symmetric gauge
\beq\lb{AAA}
\vec A(x,y)= \frac{B}{2}(y,-x).
\eeq
to find a second order equation for $\te$
\beq
\te^2 -4l_B^2 \te+ \left(\frac{2\hbar}{m\om}\right)^2=0
\eeq
where $\l_B= \sqrt{\frac{\hbar c}{eB}}$ is the magnetic length, which define the area occupied
by the Hall droplet \cite{Prange}. This can be solved to obtain
\beq
\te_{\pm} = 2l_B^2 \pm 2 \sqrt{l_B^4 -\left(\frac{\hbar}{m\om}\right)^2}
\eeq
where the condition $\l_B \geq \sqrt{\frac{\hbar}{m\om}}$ must be fulfilled. This shows another alternative
way to measure and give some interpretation for $\te$.

\section{Hall conductivities}

Actually, it known that Hall effect remains among the fascinating areas appeared
in condensed matter physics. It comes out that one can ask about the possibility
to describe such effect in terms of language. To answer this inquiry,
we will show how one can use our results to determine
explicitly
the corresponding charge and spin the conductivities.

In the spirit of the Drude model we shall now add to equation (\ref{drag})
the drag force $-m_{\theta_z}\dot{\vec{r}}/\tau$, with  $\tau$ is the relaxation time. Then we can write
\begin{eqnarray}\lb{bgbg}
m_{\theta}\ddot{\vec{r}} + {m_{\theta}\ov\tau} \dot{\vec{r}} +{\partial V(\vec{r})\ov \partial \vec{r}}
\approx \vec{F}(\vec{\theta},\vec{r},\dot{\vec{r}}).
\end{eqnarray}
In this situation  $\vec{F}(\vec{\theta},\vec{r},\dot{\vec{r}}) $ can be seen  as a perturbation to
the system described by the second order differential equation.
Taking this into consideration, we split the solution
of (\ref{bgbg}) into two parts
\beq
\dot{\vec{r}}=\dot{\vec{r}}_0+\dot{\vec{r}}_1
\eeq
where the average of the quantities $\dot{\vec{r}}_0$ and $\dot{\vec{r}}_1$ are given by
\begin{eqnarray}\lb{vec}
\langle\dot{\vec{r}}_0\rangle&=&{\tau e\ov m_{\theta}}\vec{E}\lb{vec}\\
\langle\dot{\vec{r}}_1\rangle&=&{\tau e\ov m_{\theta}}\langle \vec{F}(\vec{\theta},\vec{r}_0,\dot{\vec{r}}_0) \rangle \lb{vec2}.
\end{eqnarray}
From (\ref{force}) we obtain the relation
\beq\lb{f'}
\langle\vec{F}(\vec{\theta},\vec{r}_0,\dot{\vec{r}}_0)\rangle={e\ov c}\langle\dot{\vec{r}}_0\rangle\times \langle\vec{B}(\vec{\theta},\vec{r}_0)\rangle.
\eeq
Now using (\ref{A})  together with  (\ref{vec}) and (\ref{vec2}) to show 
\begin{eqnarray}\lb{r_1}
\langle\dot{\vec{r}}_1\rangle={e^2\tau^2\ov2m_{\theta_z}\hbar}\vec{E}\times\left({\partial\ov\partial \vec{r}}\times\left[\vec{\theta}\times\vec{E}\right]\right)
\end{eqnarray}
where the right hand side of  (\ref{r_1}) contains the volume average of ${\partial\vec{E}\ov\partial \vec{r}}$.
This can be evaluated by
%
considering
the scalar potential  $V(\vec{r})={m\omega^2\ov2}r^2$
to write ${\partial\vec{E}\ov\partial \vec{r}}=-{m\omega^2\ov e}$
and then it follows
\begin{eqnarray}\lb{r_11}
\langle\dot{\vec{r}}_1\rangle={em\tau^2\omega^2\ov2\hbar m_{\theta}}\left[\vec{\theta}\times\vec{E}\right].
\end{eqnarray}

Having established all
needed necessary material, we focus now on the derivation of
the Hall conductivities. 
For this, we begin by defining the vector of spin polarization associated
 to the noncommutative plane $\mathbb{R}^2_{\te}$
 \beq
  \vec{\xi}(\vec{\theta})=\langle\vec{\theta}\rangle
  \eeq
and absolute value ${\xi}_\theta$ lies between $0$ and $1$, such as
\beq
{\xi}_\theta={n^+-n^-\ov n^++n^-}
\eeq
where we denote by $n^\pm$ the concentrations of charge carriers with spins parallel and antiparallel to $\vec{\xi}(\vec{\theta})$, respectively, in 
$\mathbb{R}^2_{\te}$. It is convenient to introduce the total concentration
 $n=n^++n^-$  of charges carrying the electric current in $\mathbb{R}^2_{\te}$. The density matrix of the
charge carriers in the spin space can be written as 
\beq
N(\vec{\theta})={1\ov2}n\left(1+\vec{\xi}(\vec{\theta})\cdot\vec{\theta}\right).
\eeq
We choose the spin polarization vector $\vec{\xi}(\vec{\theta})$ along the $z$-direction,
i.e. $\vec{\xi}(\vec{\theta})= {\xi}_\theta \vec e_z$, and adopting the definition of the electric current
\cite{Chudnovsky} 
\beq
\vec{j}=e\langle N(\vec{\theta}) \dot{\vec{r}}\rangle=e\langle N(\vec{\theta}) (\dot{\vec{r}}_0+\dot{\vec{r}}_1)\rangle.
\eeq
It can be evaluated by using  (\ref{vec}) and  (\ref{r_11}) to write
\beq
\vec{j}=\sigma_c({\theta})\vec{E}+\sigma_s({\theta})\left(\vec{\xi}(\vec{\theta})\times\vec{\theta}\right)
\eeq
and therefore the deformed charge
and spin Hall conductivities read as
\begin{eqnarray}
\sigma_c({\theta})&=&{n e^2\tau\ov m_{\theta}}\lb{sig}\\
\sigma_s({\theta})&=&{n e\tau^2m\omega^2\ov4m_{\theta}\hbar}\lb{sig2}
\end{eqnarray}
which are clearly noncommutative parameter $\te$-dependent.  This will offer
different discussions and interpretation
in the forthcoming analysis.

Having obtained our conductivities, let us make
different discussions and offer some interpretation. These will be the subject of
 the following points:\\
$\blacktriangleright $ 
By switching off $\te$ in \eqref{sig} and \eqref{sig2}, we can easily obtain
\begin{eqnarray}
\sigma_c&=&{n e^2\tau\ov m} \lb{iscc}\\
\sigma_s&=&{n e\tau^2\omega^2\ov4\hbar}\lb{isss}
\end{eqnarray}
which can be identified to that obtained in  \cite{Chudnovsky}
where $\om^2$ has to be taken proportional to the system area.
This first connection tells us that 
are general so that  one can extract more
information about the present system by playing with the values taken by $\te$.
More detail about this point will give in the next section.\\
$\blacktriangleright$ By looking at the ratio 
\begin{eqnarray}\lb{sig''}
{\sigma_s(\te)\ov\sigma_c(\te)}={\sigma_s\ov\sigma_c}=
{m\omega^2\tau\ov4e\hbar}
\end{eqnarray}
one realizes immediately that it is  independent of the concentration of charge carriers. This  is in agreement with
the result obtained  in different microscopic models
of SHE \cite{halperin,tse}. More discussion about such point and its
relation with temperature can be found in \cite{Chudnovsky}.\\
$\blacktriangleright$ To make comparison with already published
work,
we write \eqref{sig} and \eqref{sig2}
in terms of the standard quantities $\si_c$ and $\si_s$ as
\begin{eqnarray}\lb{sig'}
\sigma_c({\theta})&=&\si_c+{n e^2\tau m\omega^2\ov4\hbar^2}\theta^2 \lb{sig'}\\
\sigma_s({\theta})&=&\sigma_s+{n e\tau^2m^2\omega^4\ov16\hbar^3}\theta^2\lb{sig222}
\end{eqnarray}
which are parabolic functions in terms of the noncommutative parameter. It is clear that
from \eqref{sig'} and \eqref{sig222}, the conductivities  can be
controlled either by adjusting the concentration $n$ of charge carriers or parameter $\theta$.
This latter can be regarded
as an external source in similar way to the magnetic field in the systems
exhibiting the quantum Hall effect~\cite{Prange}. Furthermore,
the second terms in the above equations can be interpreted  as quantum corrections to the standard
results, which consists another way to look at our findings. \\
$\blacktriangleright$ The Hall conductivities obtained
using different approach and techniques \cite{dayi,ma,basu} are linear function of $\te$ rather than quadratic as we have
in \eqref{sig'} and \eqref{sig222}. This difference is due to the  manifestation of the confining
potential taken into account in the noncommutative plane.\\
$\blacktriangleright$ Our spin conductivity can be identified to quantized ones obtained
by using other approaches. With that one can make a quantization of the noncommutative parameter
and fix some of its experimental values. Indeed on the light of the results reported in \cite{Chudnovsky}, we can
use the experimental data
on aluminum 
to give a table showing some of the $\te$
values.

\section{Interesting cases}

Having obtained the general forms of the Hall conductivities, let us show how to recover
some significant results derived in different microscopic models. Indeed,
In studying SHE Chudnovsky \cite{Chudnovsky} considered
the following Hamiltonian
in  real space
\beq\lb{ggg}
H={1\over 2m}
\left(p_{x}^{2 }+p_{y}^{2}\right) +V(\vec{r})
\eeq
where the scalar potential is defined as
\beq\lb{chpo}
V(\vec{r})={4\pi\hbar^2e^3Zn_0\ov 3m^2c^2}r^2
\eeq
with -$Ze$ and $n_0$ are the charge and the concentration of ions respectively. This model
was used to obtain similar relations to \eqref{iscc} and \eqref{isss}. Now it clear that,
these results can be derived from our approach by taking $\te=0$ and fixing the frequency as
\beq
\om= \sqrt\frac{8\pi\hbar^2e^3Zn_0}{3m^3c^2}.
\eeq
On the other hand, we can keep the noncommutative parameter and require an identification
between our Hall conductivities and those obtained by Chudnovsky. This will offer
another way to
fix $\te$ in terms of different parameters
appearing in the potential \eqref{chpo} and also
allow for establishing a link
with the experimental data.

Another interesting theory that can be derived from our results is
the Landau problem, which is describing one particle living on the plane
in the presence of an uniform magnetic field $B$. More precisely, in the limit $\theta\longrightarrow 0$
we show that (\ref{rrr})
has a strong overlapping with such problem and therefore it can be adopted to reproduce its basic features.
Then in such limit we can approximate
the effective mass (\ref{m_te}) by writing $m_\te \simeq m$
and now the Hamiltonian (\ref{rrr}) reduces
\beq\lb{weak-te}
H_{\theta\rightarrow 0}^{\sf nc}={1\ov2m}\left[ \left(p_x^{2}+p_y^{2}\right)
+m^2\omega^{2}{\left(x^{2}+y^{2}\right)+{m^2\omega^{2}\over\hbar}\te\left(yp_x-xp_y\right)}\right]
\eeq
which means that we keep only the first order term in $\te$ and drop the second one. One can see
that \eqref{weak-te}   is actually sharing some common features with the Landau problem on the plane.
To clarify this statement, let us
chose the
symmetric gauge
$A={B\ov2}(-y,x)$ to write the Hamiltonian for one charged particle living on plane and in
magnetic field as
\beq\lb{landau}
H_{\textmd{landau}}={1\ov2m}\left[ \left(p_x^{2}+p_y^{2}\right)
+\left({eB\ov2c}\right)^2{\left(x^{2}+y^{2}\right)+{eB\ov c}\left(yp_x-xp_y\right)}\right]
\eeq
Now it clear that  $H_{\textmd{landau}}$ and $H_{\theta\rightarrow 0}^{\sf nc}$
are similar and  then
 one can go from one to another  by using
the following mapping
\beq
\theta^{\textmd{landau}}={4\hbar c\ov eB}=4l_B^2,\qquad \omega={\omega_c\ov2}
\eeq
 where $l_B$ is the magnetic length and
 $\om_c=\frac{eB}{mc}$ is the cyclotron frequency.   Therefore  $\theta$ plays actually
 the role of the magnetic field $B$, which is not
 surprising because it known that
 at LLL the position coordinates become noncommuting.
This analogy will allow us to bring different results regarding $B$ straightforward
to $\te$. Indeed,
since the filling factor of the Landau levels
reads as the ratio between the density of charge carriers $\rho$
and $B$
\beq
\nu_B= \frac{\rho e c}{B}
=
\frac{l_B^2 \rho}{2\pi}.
\eeq
Then in similar way we write
\beq\lb{tenu}
\nu_\te = \frac{\te\rho}{8\pi}. 
\eeq
It gives another way to measure $\te$ in terms of the observed
quantized plateaus \cite{Prange} and one illustration will be
done soon.

Since (\ref{landau}) is the cornerstone of the
quantum Hall effect \cite{Prange}, then $H_{\theta\rightarrow 0}^{\sf nc}$  will allow us to build a theory
similar to that of the Landau problem.
Let us consider $N$ particles described by
\beq
{\cal{H}}= \sum_{i=1}^N \left(H_{\theta\rightarrow 0}^{\sf nc}\right)_i
\eeq
where $\left(H_{\theta\rightarrow 0}^{\sf nc}\right)_i$ is for one particle given in
(\ref{landau}). Starting from the above Hamiltonian,
we can construct the  Laughlin wavefunction~\cite{laughlin} for the filling factor $\nu=1$
as Slater determinant and then
we generalize to write
\beq\lb{lwave}
\Phi_{\textsf{L}}^{l} (z,\bar{z},\theta)=\prod_{i<j}(z_i-z_j)^{2l+1}\exp\left(-{1\ov \theta}\sum_i|z_i|^2\right)
\eeq
where $z=x+iy$ complex variable,  $\nu = {1\ov 2l+1}$ and $l$  is integer value.
Compared to the Laughlin states, we can interpret \eqref{lwave} as describing
Hall liquid where each droplet is occupying an area of surface $\frac{\pi}{2}\te$.
Now according to \eqref{tenu}, we can quantize $\te$
\beq
\te'_l=\frac{\te_l}{8\pi/\rho} =  \frac{1} {2l+1}
\eeq
We can go further and talk about other issues related the Laughlin wavefunctions \eqref{lwave}
like for instance fractional charges.

\section{Conclusion}

A system of two harmonic oscillators with the same masses and frequencies  were considered
on the noncommutative plane. This theoretical
model offered for us a mathematical tool to develop an approach in studying  the basic features
of the present system and in particular the Hall effect. More precisely, by considering
plane coordinates are noncommuting, we have obtained an effective  Hamiltonian involving
coupling term that was interpreted as spin-orbit coupling where
the noncommutative parameter $\te$ is identified to the spin.
Moreover, we have obtained a noncommutative mass $m_\te$, which was linked to mass of
a relativistic particle moving with velocity $v$. This connection
allowed for another possibility to give some measurement for $\te$ and thus
an experimental proof. Using the algebraic method, we have
determined explicitly the solution of the energy spectrum in terms of $\te$.
By requiring that $\te=0$, we have seen that the standard results can be recovered easily.

The symmetry was also taken part of our investigation. Indeed,
in terms of the quadratic observables, we have realized two algebras: $su(2)$
and $su(1,1)$. For each algebra, we have discussed the corresponding unitary irreducible
representations and their relations to  our eigenvalues. Later on, we have analyzed three
interesting cases in filling the shells with fermions:
week, strong and generic coupling. The first case showed a manifestation of
$su(2)$ where the level degeneracies were fixed
and the filling of shells was done.  
While in the second case, we have found an infinite degeneracy and
our eigenstates are ladder states for the discrete series representation labeled by $-k + 1/2$.

Subsequently, we have shown that it is possible to derive
an effective Lorentz force as function of $\te$ similar to
that acting on one particle living on plane in the presence of magnetic field $B$. Therefore, the gauge field
$\te$-dependent
was established and its relation to that corresponding to $B$
was discussed. Indeed, after identification with symmetric gauge, we have obtained
a second order equation for $\te$, which was solved to
express $\te$  in terms of the magnetic length
$\l_B$. This allowed for another way to describe the noncommutative
parameter and then linked with experimental data.

By adding a drag term to the obtained second order differential equation for the position $\vec r$,
we were able to talk about the Hall conductivities
exhibited by our system. Indeed, such equation allowed us to get the corresponding velocity that was used
to calculate the electric current according to definition adopted in~\cite{Chudnovsky}. Doing this processing
to finally find the charge and spin Hall conductivities as quadratic functions of $\te$.   More precisely, these
two conductivities showed extra terms in addition to the standard one, which were regarded as quantum
corrections.

Furthermore, we have analyzed interesting limiting cases, which were lead to recover already published works. Indeed,
by fixing our frequency as function of different physical parameters and switching off $\te$, we have recovered
the Chudnovsky results.
By considering the limit $\te\lga 0$, our noncommutative Hamiltonian was reduced to another one, which
has some common features with the Landau problem. This allowed us to make a mapping between both
and therefore establish a bridge to Laughlin states where $\te$ was quantized in terms of filling factor.

\section*{Acknowledgments}

The generous support provided by the Saudi Center for Theoretical Physics (SCTP)
is highly appreciated by all authors. AJ acknowledges partial support
by King Faisal University.

\end{document}